# Quantitative Parametric Mapping of Tissues Properties from Standard Magnetic Resonance Imaging Enabled by Deep Learning

Yan Wu[1,2,+], Yajun Ma[3,+], Youngwook Kee[2], Nataliya Kovalchuk[1], Dante Capaldi[1], Hongyi Ren[1], Steven Hancock[1], Eric Chang[3], Marcus Alley[2], John Pauly[2], Jiang Du[3,*], Shreyas Vasanawala[2,*], Lei Xing[1,*]

[1] Radiation Oncology Department, Stanford University, 875 Blake Wilbur Drive, G204, Stanford, California 94305
[2] Radiology Department, Stanford University, 300 Pasteur Drive, Stanford, California 94305
[3] Radiology Department, University of California San Diego, 9500 Gilman Drive, #0888, La Jolla, California 92093
[*] corresponding author
[+] equal contribution

**Summary**

**Magnetic resonance imaging (MRI) offers superior soft tissue contrast and is widely used in biomedicine. However, conventional MRI is not quantitative, which presents a bottleneck in image analysis and digital healthcare. Typically, additional scans are required to disentangle the effect of multiple parameters of MR and extract quantitative tissue properties. Here we investigate a data-driven strategy $Q^2MRI$ (Qualitative and Quantitative MRI) to derive quantitative parametric maps from standard MR images without additional data acquisition. By taking advantage of the interdependency between various MRI parametric maps buried in training data, the proposed deep learning strategy enables accurate prediction of tissue relaxation properties as well as other biophysical and biochemical characteristics from a single or a few images with conventional $T_1/T_2$ weighting. Superior performance has been achieved in quantitative MR imaging of the knee and liver. $Q^2MRI$ promises to provide a powerful tool for a variety of biomedical applications and facilitate the next generation of digital medicine.**

**Introduction**

MRI exploits the magnetic properties of tissues and provides noninvasive imaging of the human body for a wide variety of clinical applications [1]. While adopted ubiquitously, current MRI techniques cannot directly evaluate inherent physical properties of tissue (e.g., $T_1$ or $T_2$ relaxation times), which presents a bottleneck in image analysis and precision medicine [2]. The signal intensity of an MR image depends not only on tissue properties themselves, but also acquisition parameters (e.g., time of repetition, echo time, flip angle) and condition of the MR system being used (e.g., homogeneities of the radiofrequency field $B_1$ and main magnetic field $B_0$, uniformity of receiver coil sensitivity). In general, the MRI parameter space can be divided into three subspaces, corresponding to the tissue relaxation properties, the acquisition parameters, and the system conditions (Fig. 1a). Conventionally, an MR image is acquired by using a pre-selected pulse sequence with specific imaging parameter setting, and the resulting image is qualitative in nature, with its signal intensity being a function of both tissue properties and other factors in the MRI parameter space. In order to derive a quantitative parametric map that reflects only a tissue property, one has to perform multiple samplings in the acquisition parameter subspace and then solve the system equations.

In past decades, extensive research efforts have been devoted to finding effective means to rapidly explore the multi-parametric space for quantitative MR imaging. A conventional approach quantifies a single tissue relaxation parameter by extracting the exponential decay/regrowth rate from multiple measurements (Fig. 1b) [3-7]. This strategy requires a substantially long scan time and is prone to system imperfections. Magnetic/radiofrequency field maps are sometimes measured to compensate for inaccuracy caused by field inhomogeneities [8-10]. Moreover, advanced image reconstruction techniques, such as compressed sensing and low rank processing, can be incorporated into quantitative MRI [11-14]. Then multi-parametric mapping methods have been developed, such as MR fingerprinting, MAGiC, multi-tasking, and EPTI [15-23]. With the help of advanced parameter fitting techniques, various tissue relaxation properties and field maps can be simultaneously estimated from the same input images (Fig. 1c). However, special pulse sequences are required. Recently, deep learning [24] has been employed to accelerate the data acquisition or post-processing in quantitative parametric mapping [25, 26] (as well as in other medical imaging applications [27]). Despite these remarkable speed-ups, additional scans are always required, hindering widespread adoption of quantitative MRI in clinical practice.

Here, we report a data-driven solution for quantitative measurement of inherent tissue properties with automatic compensation for system imperfections. Fundamentally different from previous approaches, the $Q^2MRI$ technique leverages deep learning to exploit the inherent relationship within and across MRI parameter subspaces to derive quantitative parametric maps from standard MRI acquisitions. In our process, a significant portion of the information needed for parametric mapping is obtained from training data acquired using different pulse sequences. As a result, it becomes possible to predict tissue relaxation properties and field maps from a routine MRI examination without additional data acquisition or change of clinical workflow (Fig. 1d). Furthermore, biophysical and biochemical parametric maps — traditionally obtained with the involvement of confounding models (e.g., quantitative susceptibility map and proton density fat fraction map [28, 29]) — can be derived from the same framework without further processing. In this sense, the proposed strategy goes beyond the scope of multi-parametric mapping for relaxation properties. Applications of $Q^2MRI$ to the knee and liver show that the proposed technique provides an accurate and efficient way for quantitative MR imaging.

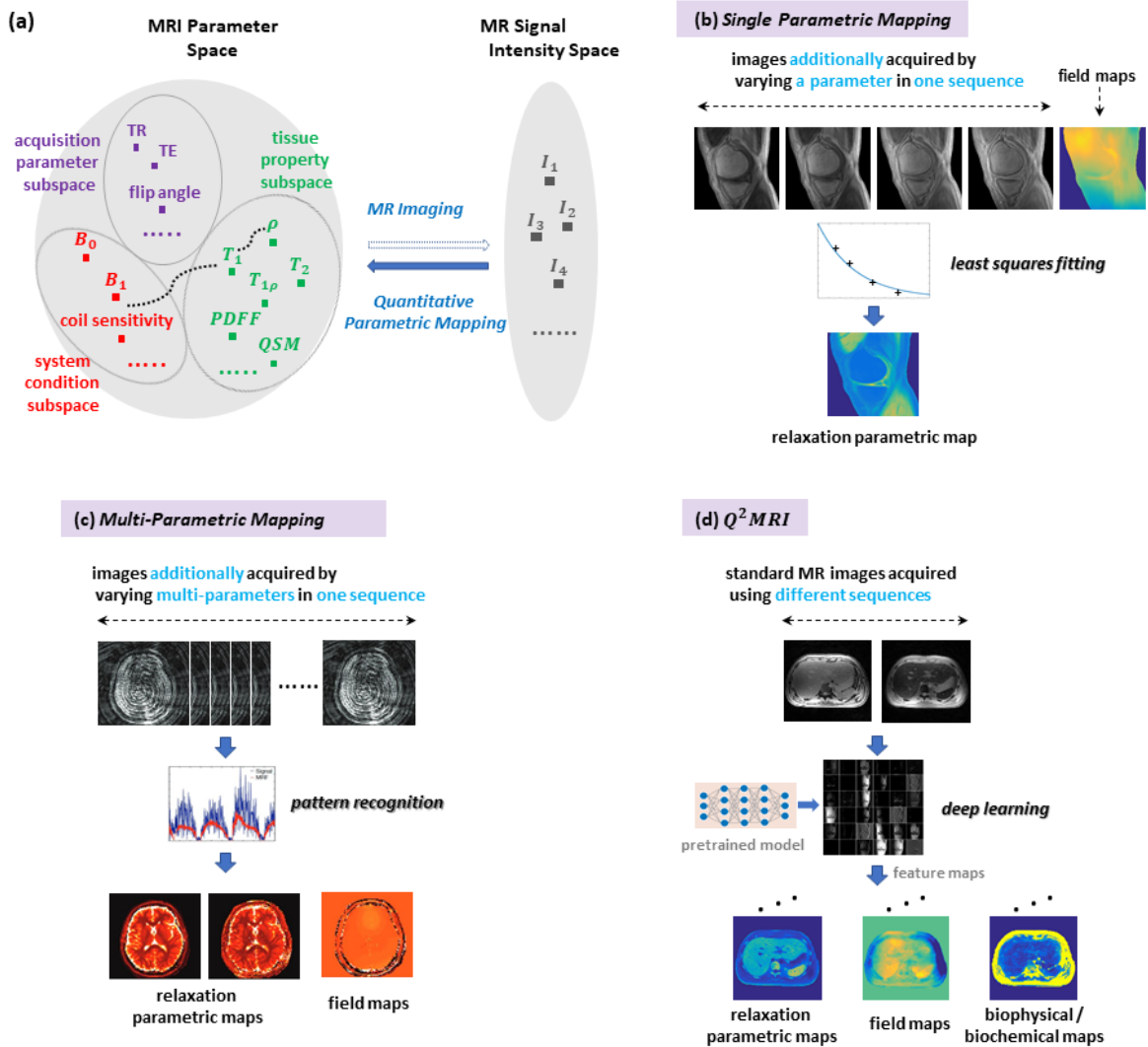

**Figure 1.** An illustration of conventional and proposed quantitative parametric mapping strategies. (a) The MRI parameter space consists of three subspaces, representing tissue properties, acquisition parameters, and system conditions, respectively. Quantitative parametric mapping can be achieved by sampling the acquisition parameter subspace and then solving the inverse problem from multiple measurements. (b) In single parametric mapping, a quantitative tissue relaxation parametric map (e.g., $T_1$) is extracted from several images acquired using the same pulse sequence but different acquisition parameters (e.g., flip angle), and the result is further compensated by field map (e.g., $B_1$). Here, the field map is specifically measured, and parameter quantification is typically implemented using least squares fitting. (c) In multi-parametric mapping (e.g., MR fingerprinting), a series of images is acquired using a specially designed pulse sequence with variations in multiple acquisition parameters (e.g., repetition time, echo time, and flip angle); subsequently, tissue relaxation properties ($T_1$, $T_2$) and field maps ($B_0$, $B_1$) are simultaneously estimated using advanced parameter fitting techniques (e.g., pattern recognition). (d) In the proposed $Q^2MRI$ method, tissue relaxation properties (e.g., $T_1$, $T_2$) and field maps (e.g., $B_0$, $B_1$) are derived from one or a few MR images with conventional $T_1/T_2$ weighting, which are obtainable in clinical practice. Other biophysical and biochemical parametric maps (e.g., proton density fat fraction map) can also be derived in the same deep learning framework without further processing. The essence of the proposed method is to use deep learning to exploit *a priori* information on various parameters in the MRI parameter space (within/across different subspaces), mitigating the need to make actual measurements as is required by parameter quantification.

## Results

In this study, $Q^2MRI$ was validated in $T_1$ and $T_2^*$ mapping of the knee, and in $R_2^*$ mapping of the liver, where deep neural networks were used to provide end-to-end mappings. In the knee, $T_1$, $\rho$, and $B_1$ maps were predicted from single $T_1$- weighted images, and $T_2^*$ maps were derived from single $T_1$- and $T_2^*$-weighted images. In the liver, $R_2^*$, $B_0$, and proton density fat fraction ($PDFF$) maps were estimated from pairs of in-phase and out-of-phase images (Fig. 2).

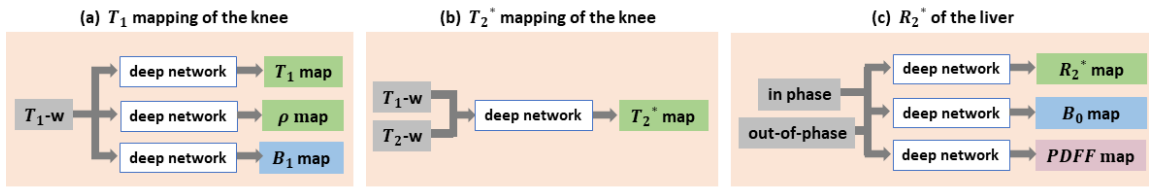

**Figure 2.** In $Q^2MRI$, deep learning models are trained to provide end-to-end mapping from single images to quantitative parametric maps and field maps.

## $T_1$ and $T_2^*$ Mapping of the Knee

For $T_1$ mapping of the knee, $T_1$, $\rho$, and $B_1$ maps were predicted from single $T_1$-weighted images. Here, the ground truth $T_1$ map was extracted from four $T_1$-weighted images (acquired using variable flip angles of 5°, 10°, 20°, and 30°, respectively) via least squares fitting, and further corrected by the measured $B_1$ map; $\rho$ map was calculated from a $T_1$-weighted image and $T_1$ map. The input image was one of the four $T_1$-weighted images (acquired using a specific flip angle of 5°, 10°, 20°, or 30°). A total of 1224 two-dimensional (2D) images from 59 subjects (including 50 osteoarthritis patients and 9 healthy volunteers) were used for training and testing, with six-fold cross-validation applied.

High fidelity mapping was achieved consistently across patients. A representative case is shown in Fig. 3. In the resultant $T_1$ maps, compensation for $B_1$ inhomogeneity was automatically achieved without use of a measured $B_1$ map. It is intriguing that the accurately predicted $B_1$ map was implicitly incorporated into the $T_1$ mapping models, which mitigated the need for actual $B_1$ measurement. On the individual pixel level, high correlation coefficients, low $l_1$ errors and high SSIM were obtained for various maps that were derived from different input images (Table 1). We note that slightly higher accuracy was achieved from input images that had been acquired using a flip angle of 20° or 30°, where $T_1$ effect was more prominent. Within the regions of interest (ROIs) in femur, tibia, and patella cartilage (which were manually segmented), averaged $T_1$ was calculated; the Bland-Altman plots are shown in Fig. 4.

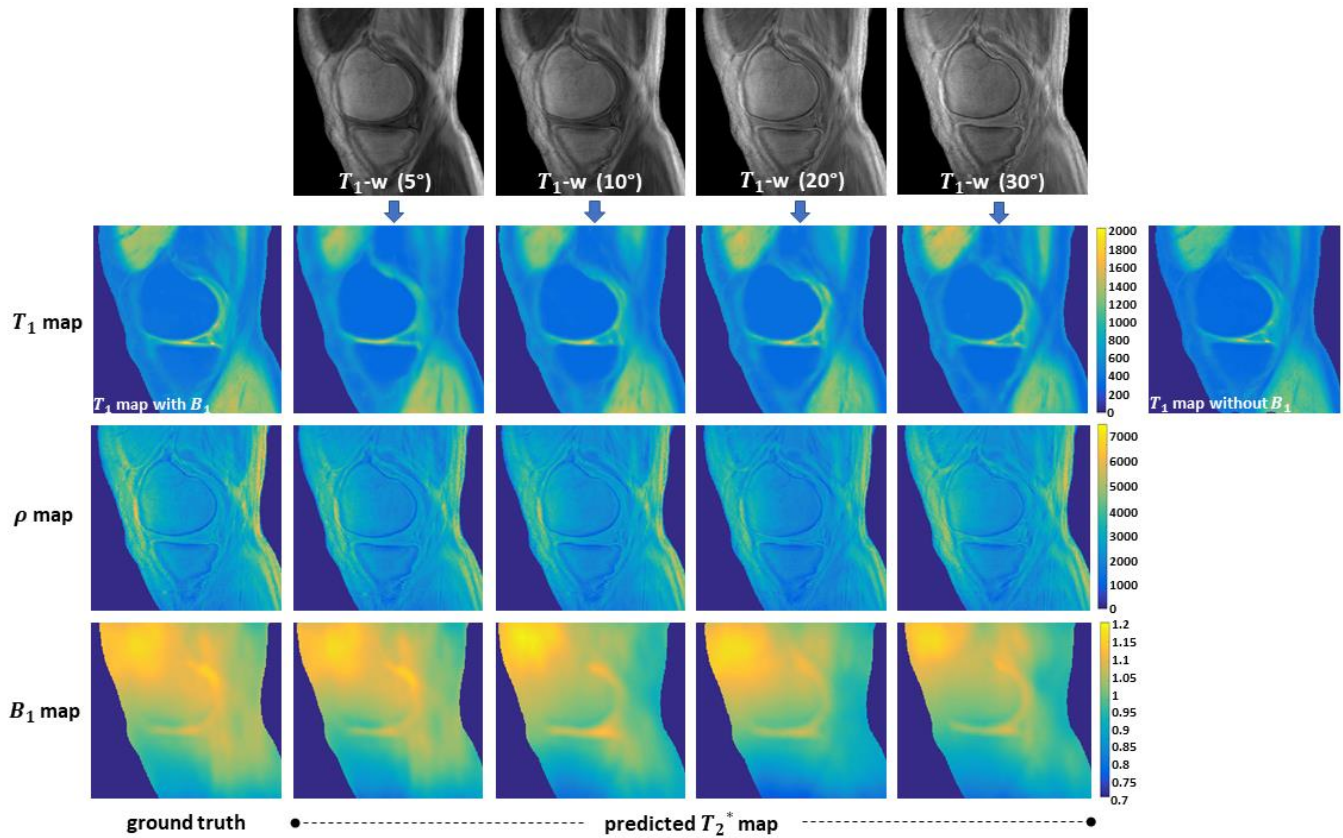

**Figure 3.** Prediction of $T_1$, $\rho$, and $B_1$ maps of the knee from single $T_1$-weighted images acquired using a specific flip angle (5°, 10°, 20°, and 30° respectively). The quantitative parametric maps and field maps predicted from different input images are all very close to the ground truth maps. Particularly in the resultant $T_1$ maps, compensation for $B_1$ inhomogeneity is automatically achieved without use of the measured $B_1$ map.

| Output map | Input image | $T_1$ w (5°) | $T_1$ w (10°) | $T_1$ w (20°) | $T_1$ w (30°) |
|---|---|---|---|---|---|
| $T_1$ map | $l_1$ Error | 0.1026 ± 0.0311 | 0.0986 ± 0.0311 | 0.0952 ± 0.0283 | 0.0937 ± 0.0304 |
| | Correlation | 0.9681 ± 0.0276 | 0.9749 ± 0.0185 | 0.9778 ± 0.0170 | 0.9786 ± 0.0160 |
| | SSIM | 0.7743 ± 0.0430 | 0.7862 ± 0.0338 | 0.8151 ± 0.0303 | 0.8225 ± 0.0262 |
| $\rho$ map | $l_1$ Error | 0.0685 ± 0.0098 | 0.0566 ± 0.0076 | 0.0635 ± 0.0064 | 0.0642 ± 0.0080 |
| | Correlation | 0.9914 ± 0.0036 | 0.9937 ± 0.0033 | 0.9922 ± 0.0036 | 0.9919 ± 0.0038 |
| | SSIM | 0.8595 ± 0.0346 | 0.9059 ± 0.0263 | 0.8910 ± 0.0263 | 0.9173 ± 0.0151 |
| $B_1$ map | $l_1$ Error | 0.0500 ± 0.0210 | 0.0588 ± 0.0253 | 0.0723 ± 0.0302 | 0.0561 ± 0.0273 |
| | Correlation | 0.9954 ± 0.0040 | 0.9947 ± 0.0036 | 0.9917 ± 0.0044 | 0.9956 ± 0.0038 |
| | SSIM | 0.9979 ± 0.0013 | 0.9974 ± 0.0017 | 0.9962 ± 0.0028 | 0.9976 ± 0.0020 |

**Table 1.** The quantitative results for $T_1$, $\rho$, and $B_1$ mapping of the knee on a pixel basis (mean ± std)

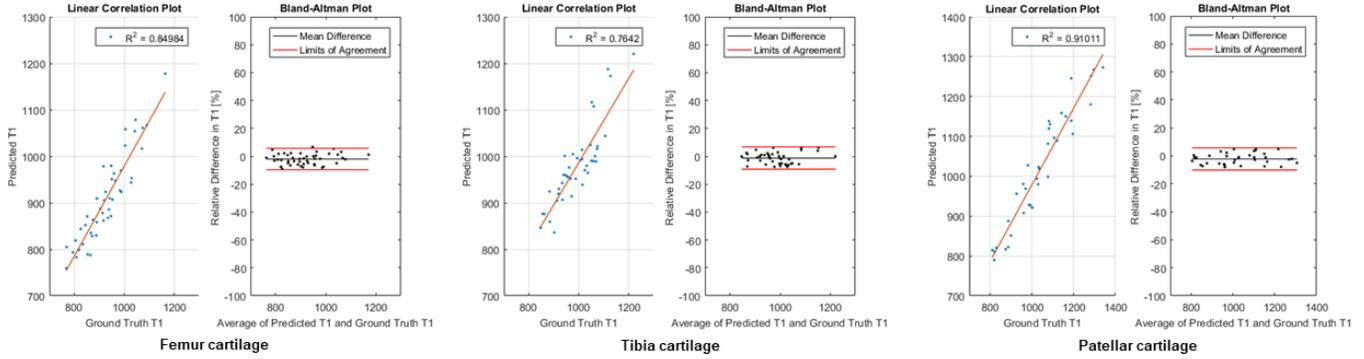

**Figure 4**. The Bland-Altman plots for $T_1$, where every dot corresponds to the averaged $T_1$ within an ROI (in femur, tibia, or patella cartilage) per subject.

$T_1$ maps predicted from single input images closely resembled the maps derived from two $T_1$-weighted images. In Fig. 5, $T_1$ maps derived from one image acquired with a flip angle of 20° or 30° were highly similar to the map predicted from both images. On the pixel level, correlation coefficients were 0.9778, 0.9786, and 0.9836 when input images were acquired using a flip angle of 20°, 30°, or both, and $l_1$ errors were 0.0952, 0.0937, and 0.0567, respectively.

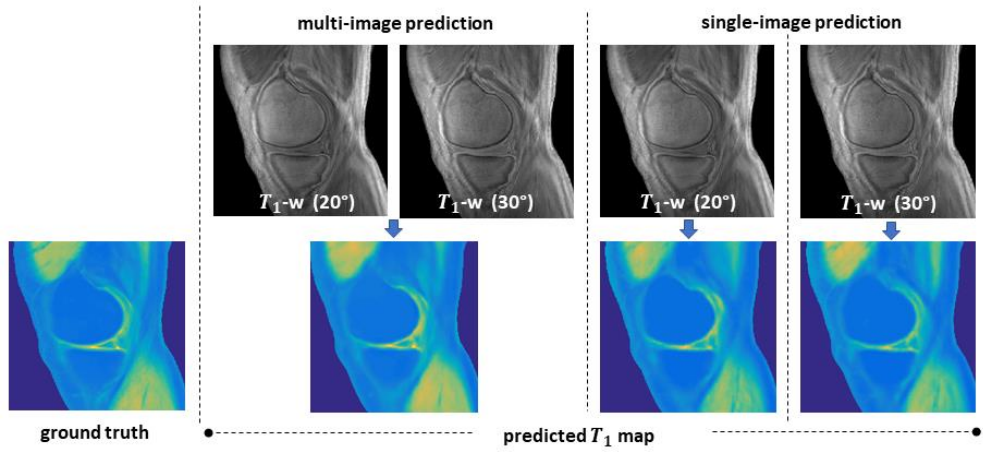

**Figure 5.** Comparison of $T_1$ maps of the knee predicted from single input images (acquired using a flip angle of 20° or 30°) with $T_1$ map derived from both input images. The single input predictions are highly consistent with the dual-input prediction.

It is remarkable that $T_1$ maps were accurately predicted from a single input image. We believe that the underlying reason for these results is that deep learning takes advantage of the correlation between various quantitative maps and integrates this information into the parametric mapping models as *a priori* knowledge. To validate this argument, we trained deep neural networks to predict $T_1$ map from $\rho$ map and vice versa. An example is shown in Fig. 6. In the mapping from $\rho$ map to $T_1$ map, the correlation coefficient and $l_1$ error were 0.95±0.03 and 0.09±0.01, respectively; in the inverse mapping, the correlation coefficient and $l_1$ error were 0.94±0.02 and 0.13±0.02. Recovery of a high-resolution $\rho$ map from a $T_1$ map is more challenging because the ground truth $T_1$ map was slightly smoothed during measurement. This experiment supports the notion that the deep learning models successfully exploit the high correlation between $T_1$ and $\rho$ maps, which is believed to serve as *a priori* knowledge and thus effectively reduce the number of input images required for quantitative parametric mapping.

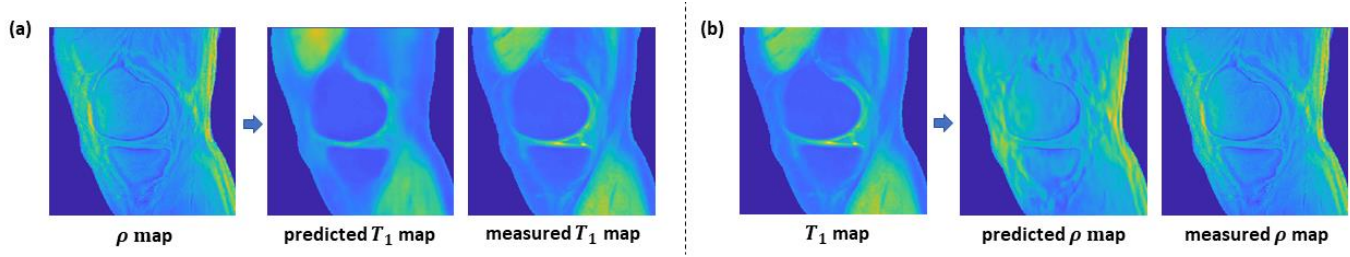

**Figure 6**. Exploiting the correlation between various tissue parametric maps. (a) $T_1$ map is predicted from $\rho$ map, confirming that high correlation between the two parametric maps has been deciphered by deep learning. (b) $\rho$ map is predicted from $T_1$ map. It is more challenging to recover high-resolution $\rho$ map from $T_1$ map because $T_1$ map was slightly smoothed during actual measurement. These experiments indicate that high correlation between $T_1$ map and $\rho$ map has been exploited by deep learning.

$T_2^*$ maps of the knee were predicted from pairs of $T_2^*$-weighted and $T_1$-weighted images. Notice that no additional scan is required because multi-contrast images have already been acquired in standard clinical MRI. Here, the ground truth $T_2^*$ map was measured from six $T_2^*$-weighted multi-echo images acquired with echo times (TEs) of 0.032, 4.4, 8.8, 13.2, 17.6, and 22 ms, respectively; input images included one of the $T_2^*$-weighted images (acquired using a specific TE of 4.4, 8.8, 13.2, or 17.6 ms) as well as a $T_1$-weighted image (obtained with a flip angle of 20°). A total of 1224 2D images from 59 subjects (including 50 osteoarthritis patients and 9 healthy volunteers) was used for training and testing, with six-fold cross-validation applied.

Excellent agreement was found between the $T_2^*$ maps predicted by deep learning models and the ground truth. A representative case is shown in Fig. 7a. High accuracy was achieved with $T_2^*$-weighted images predicted using different TEs. The Bland-Altman plots for averaged $T_2^*$ within the femur, tibia, and patella cartilage ROIs are shown in Fig. 7b.

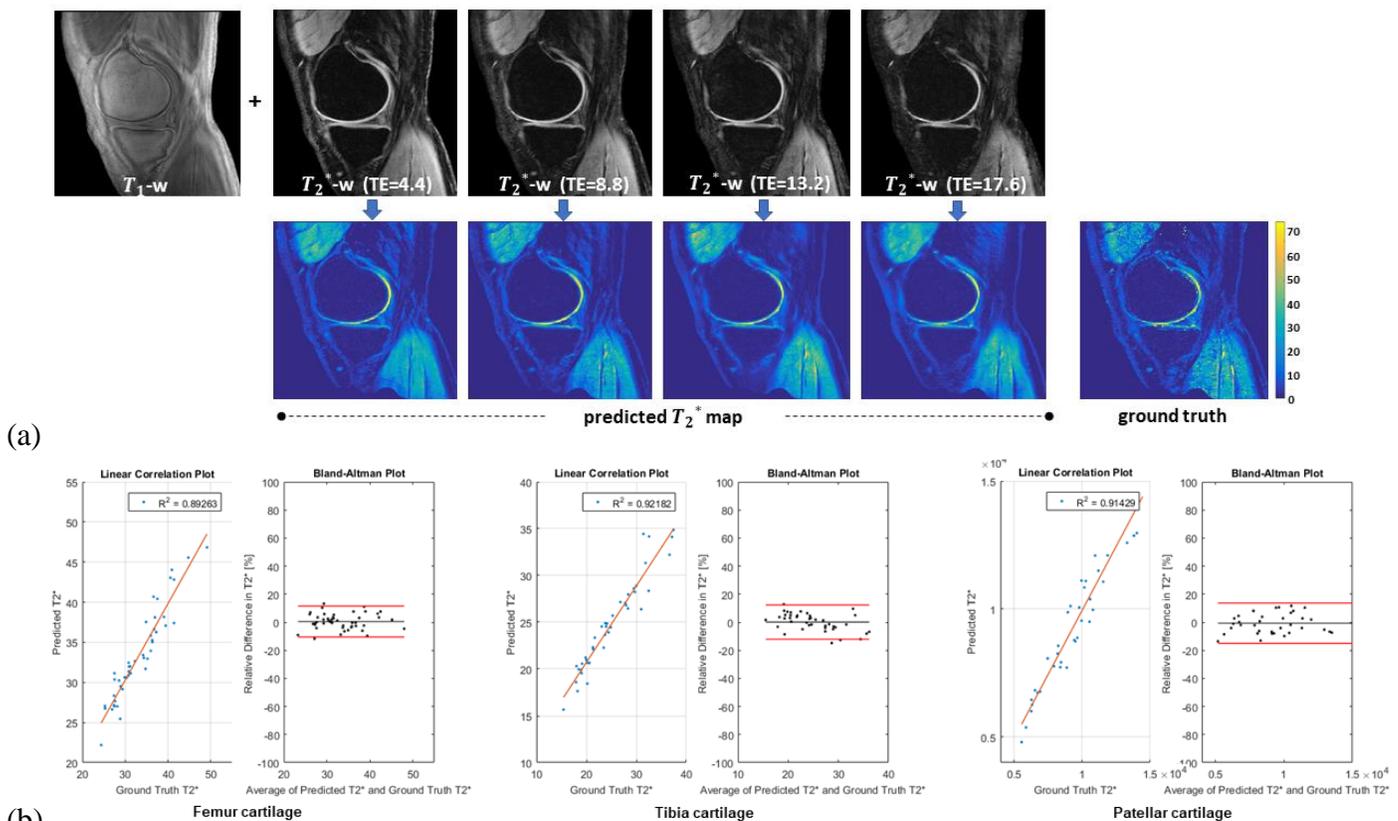

**Figure 7**. Prediction of $T_2^*$ maps of the knee from $T_2^*$-weighted images. (a) $T_2^*$ maps are predicted from single $T_2^*$-weighted images (acquired with a specific TE) and a $T_1$-weighted image (obtained using a flip angle of 20°). The $T_2^*$ maps predicted from different input images all have high fidelity to the ground truth maps displayed in the rightmost column. (b) The Bland-Altman plot for $T_2^*$, where every dot corresponds to averaged $T_2^*$ within an ROI (in femur, tibia, or patella cartilage) for a subject.

### $R_2^*$ *Mapping of the Liver*

In the liver study, $R_2^*$, $B_0$, and $PDFF$ (proton density fat fraction) maps of the liver were predicted from pairs of in-phase and out-of-phase, $T_2^*$-weighted images. Here, the ground truth $R_2^*$, $B_0$, and $PDFF$ maps were extracted from six multi-echo images using a confounder-corrected model [30]; input images included an in-phase image and an out-of-phase image obtained using a TE of 1.048 ms or 4.084 ms. A total of 1224 2D images (from 26 patients with iron overload) was used for training and testing, where leave-one-out cross-validation was applied.

The resultant maps showed strong correlations to the ground truth. A representative case is shown in Fig. 8a and additional examples of $R_2^*$ maps with various iron overload levels are displayed in Fig. 8b. Within the ROI (i.e., whole liver automatically segmented using the region growing algorithm [31]), the averaged $R_2^*$ was calculated on every slice; the corresponding Bland-Altman plot is shown in Fig. 8c.

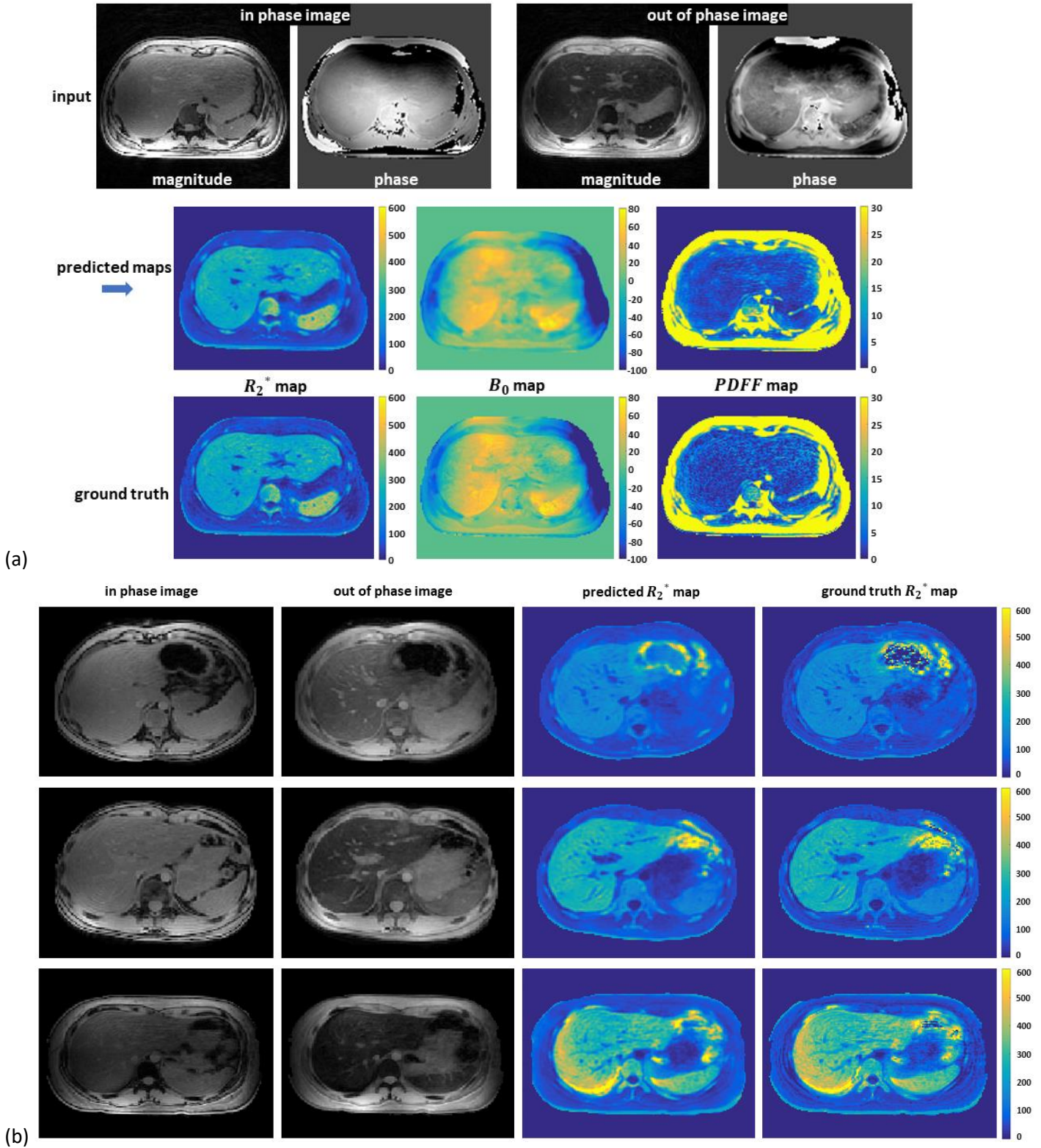

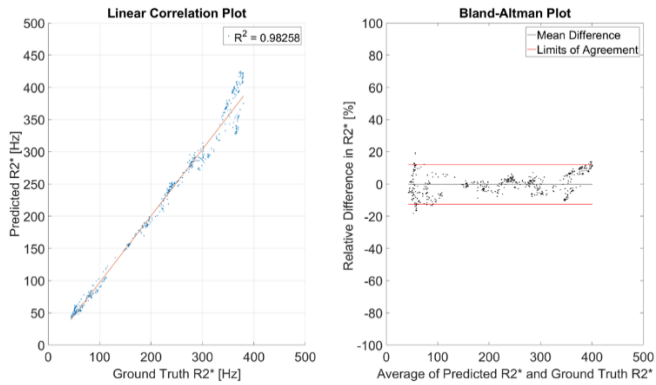
(c)

**Figure 8**. Prediction of $R_2^*$, $B_0$, and $PDFF$ maps of the liver from a pair of in-phase and out-of-phase images. (a) From the magnitude/phase of two $T_2^*$-weighted images (acquired using TEs of 1.048 ms and 4.084 ms), $R_2^*$, $B_0$, and $PDFF$ maps are predicted (middle row) with high fidelity to the ground truth (bottom row). (b) $R_2^*$ map from three patients with mild, median, and severe iron overload. Each row corresponds to one patient, and high $R_2^*$ values indicate heavy iron overload. The left two columns are input images (magnitude of in-phase/out-of-phase images), while the third and fourth columns represent the predicted and ground truth $R_2^*$ maps, respectively. (c) The Bland-Altman plot for $R_2^*$, where every dot corresponds to averaged $R_2^*$ within the ROI (i.e., whole liver) in a 2D image.

## Methods

We employed self-attention convolutional neural networks to provide end-to-end mapping from a single or very few MR images with conventional $T_1/T_2$ weighting to the corresponding quantitative parametric maps and field maps. The network has a unique shortcut connection pattern, where relatively dense local shortcuts forward feature maps to all the convolutional blocks in the same hierarchical level, and global shortcuts connect the encoder path and the decoder path [32]. An attention mechanism [33-37] is integrated into the deep neural network to make efficient use of image-wide/non-local information. In every convolutional block, a self-attention layer is combined with the convolutional layer, where the former extracts non-local knowledge (e.g., statistical distribution of tissue relaxation properties) and the latter acquires local information (e.g., smoothness of parametric map or field map) [38]. More details are given in the Appendix (A.3).

The knee images were acquired on two 3T GE MR750 scanners (GE Healthcare Technologies, Milwaukee, WI) with Institutional Review Board (IRB) approval [39, 40]. For every subject, the following images were obtained: four $T_1$-weighted images (with a matrix size of 256×256×36) acquired using flip angles of 5°, 10°, 20°, and 30°; six $T_2^*$-weighted images (of the same matrix size) acquired using TEs of 0.032, 4.4, 8.8, 13.2, 17.6, and 22 ms; and a $B_1$ map (with a matrix size of 128×128×18) measured using an actual flip angle sequence. Acquisitions for the $T_1$-weighted images, $T_2^*$-weighted images, and $B_1$ map took 9 min 28 sec, 4 min 57 sec, and 3 min 40 sec, respectively. More details are provided in the Appendix (A.1).

For $T_1$ mapping of the knee, we trained separate deep neural networks for the prediction of $T_1$, $\rho$, and $B_1$ maps from a single $T_1$-weighted image acquired using a specific flip angle (i.e., 5°, 10°, 20°, or 30°). Notice that we were able to predict the $T_1$ map without incorporating $B_1$ map as an input while expecting automatic compensation for $B_1$ inhomogeneity in the resultant $T_1$ map.

In $T_2^*$ mapping of the knee, two images with different types of weighting are used as the input because $T_2^*$ decay is a more complicated phenomenon (which could be better described by multi-component model [41]). Using a single $T_2^*$-weighted image would pose difficulty with TE selection — a long TE leads to signal loss in tissues with ultrashort $T_2$ relaxation times, but an ultrashort TE is unable to differentiate tissues with long $T_2$ relaxation times. Incorporating a $T_1$-weighted image into the input is a natural choice due to its wide availability in routine MRI examinations.

For $T_1$ and $T_2^*$ mapping of the knee, a loss function defined as $loss = l_1 + \lambda * (1 - SSIM)$ was employed [42]. Here, $l_1$ is responsible for minimizing uniform biases, and $SSIM$ (structural similarity index) preserves local structure and high-frequency contrast. λ was empirically chosen as 5. The network parameters were updated using the Adam algorithm [43] with $\alpha$ of 0.001, $\beta_1$ of 0.89, $\beta_2$ of 0.89, and $\epsilon$ of $10^{-8}$.

In the liver study, we retrospectively collected free-breathing abdominal MR images acquired on three 3T GE MR750 scanners with IRB approval [44]. For every subject, six $T_2^*$-weighted images were acquired using a multi-echo sequence with TEs of 0.036, 1.048, 2.060, 3.072, 4.084, and 5.096 ms, respectively. The in-plane resolution was 180×180, and the slice number per subject varied from 32 to 90 (leading to a scan time between 2 min 54 sec and 6 min 30 sec).

In $R_2^*$ mapping of the liver, we trained deep neural network models to predict $R_2^*$, $B_0$, and $PDFF$ maps from a pair of in-phase and out-of-phase images. Both phase and magnitude images were used for $B_0$ mapping and $PDFF$ mapping, and magnitude image was employed for $R_2^*$ mapping. In training, $l_1$ loss was adopted for $R_2^*$ and $B_0$ mapping, and the mixed $l_1$-$l_{SSIM}$ loss was used for $PDFF$ mapping.

**Discussion and Conclusions**

Derivation of quantitative tissue relaxation properties from standard MR imaging is known to be an ill-posed problem because of the highly complex nature of proton processing and its subtle dependence on not only the molecular structures themselves but also the surrounding microscopic and macroscopic environments. To understand the problem comprehensively and accurately derive mapping information, MRI signal equations corresponding to several parameters need to be solved. Previously, multiple measurements taken using the same pulse sequence have been performed for $T_1$/$T_2$ mapping. In $Q^2MRI$, images acquired using different pulse sequences are used to separate the influence of numerous contributing factors. Using the relationship between quantitative parametric maps and input images learned in training, the deep learning models can predict tissue parametric maps from a single or very few MR images with conventional $T_1/T_2$ weighting. Therefore, $Q^2MRI$ has potential to transform standard clinical MRI from qualitative imaging to quantitative imaging. Because $Q^2MRI$ is a data-driven method, its accuracy is not degraded by simplified or inaccurate model assumptions in the way that model-based methods are.

The superior capability of the proposed strategy is attributed to the powerful abilities of deep learning to integrate a variety of prior information into one comprehensive parametric mapping model. In conventional approaches, combining *a priori* knowledge is rather difficult. For example, estimating $B_1$ map from multi-contrast images was accomplished based on a physical model [7, 8, 18, 19] or a statistical model that assumed the distribution of $T_1$ in different tissues [45, 46] or a linear relationship between $1/\rho$ and $1/T_1$ in brain tissue [47]. However, there was no model that combined *a priori* knowledge in both aspects. On the contrary, deep learning learns different types of *a priori* knowledge and incorporates them into a single model, making it possible to predict the $B_1$ map from a single $T_1$-weighted image with high accuracy.

The deep learning-based modeling provides a way to exploit and utilize the interdependency between various parametric maps. According to quantum statistics [48], various tissue relaxation properties are related (Appendix A.6); and the joint distribution of these random variables is clustered [49]. While $Q^2MRI$ implicitly exploits this interdependency, additional spatial constraint is posed -- the influence of relevant voxels is taken into consideration, via the attention mechanism, for the quantification of tissue parameter at a certain voxel. This improves the robustness of prediction compared to voxel-based parameter fitting approaches. With the incorporation of physical, statistical, and spatial information, the interdependency between various parametric maps and images (e.g., $T_1$ map and $\rho$ map, $B_1$ map and $T_1$-weighted image) are deciphered and automatically encoded into the parametric mapping models, resulting in significantly reduced need for input data.

Using $Q^2MRI$, a variety of tissue properties can be predicted from very few images with conventional $T_1/T_2$ weighting. First, derivation of quantitative parametric maps from MR images without the corresponding weighting is possible (e.g., predicting $T_{1\rho}$ map from conventional MR images without $T_{1\rho}$ weighting). When the physical, statistical, and spatial relationships between $T_1$, $T_2$, and $T_{1\rho}$ are exploited, the $T_{1\rho}$ map can be estimated from $T_1$ and $T_2$ maps (which is similar to the prediction of $T_1$ map from $\rho$ map). Since $T_1$ and $T_2$ maps are predictable from few MR images with conventional $T_1/T_2$ weighting (as demonstrated in the study), $T_{1\rho}$ map is likely to be derived from the same input images without $T_{1\rho}$ weighting. This has been partially validated in a previous study, where a $T_{1\rho}$-weighted image was accurately predicted from the combination of a $T_1$-weighted image and a $T_2^*$-weighted image [50]. More generally, a variety of biophysical and biochemical parametric maps can be derived using the same strategy, where the influences of contributive factors are separated from input images and combined in a way that mimics the physical model used to extract the ground truth map. For example, in $PDFF$ mapping, $T_1$-related bias, $T_2^*$ decay, and spectral complexity of fat are believed to be implicitly extracted and used to form the confounder model (which combines least squares fitting of complex-valued source images with multi-peak fat modeling [30]).

$Q^2MRI$ is a general quantitative parametric mapping framework. The input images can be acquired using a flexible imaging protocol. For instance, $T_1$ map can be predicted from a $T_1$-weighted image acquired using a flexible flip angle. Similarly, when a slightly different imaging protocol is used to provide input images with $T_1$ and $T_2^*$ weighting, the parametric mapping models can simply be updated with transfer learning [51]. A scanner-independent model can be trained with data acquired from a collection of MRI scanners, as demonstrated in our study. The quantitative parametric mapping strategy is not confined to the proposed deep learning technique. Alternative network architectures could be employed for this task, and with the rapid advancements in artificial intelligence, further performance improvements and functionality enhancements can be expected.

$Q^2MRI$ also provides an opportunity to take full advantage of the versatility of MRI contrast. While MRI can offer versatile soft tissue contrasts to meet different clinical demands, this unique feature of MRI has never been fully utilized in practice because of the significant data acquisition overhead associated with each change of acquisition parameters. The proposed parametric mapping framework presents a simple solution that fits into a routine MRI examination workflow: after tissue properties have been estimated using $Q^2MRI$, MRI contrast can be retrospectively tuned with the application of Bloch equations. An example of changing contrast in $T_1$-weighted images (with alternative flip angles) is demonstrated in the Appendix (A.4). With the application of various pulse sequences on multi-parametric maps, a wide spectrum of image contrasts can be retrospectively obtained. In fact, $Q^2MRI$ also provides theoretical support to MRI contrast translation (i.e., direct mapping from one image contrast to another without the involvement of quantitative parametric maps). For example, predicting contrast-enhanced MRI from non-contrast enhanced images can be reframed as multi-step processing that includes estimating the $T_1$ map, changing the $T_1$ value of blood, then formulating the contrast-enhanced image based on the modified $T_1$ map. With the feasibility of parametric mapping now confirmed by $Q^2MRI$, deep learning-based MRI contrast translation has been conceptually substantiated [52, 53].

In summary, a new paradigm of data-driven $Q^2MRI$ strategy is presented for quantitative tissue parametric mapping. A significant practical benefit of $Q^2MRI$ is that no additional scans are required for generating quantitative parametric maps. The extraordinary capability of deep learning in deriving qualitative MR parametric maps from a single or very few MR images with conventional $T_1/T_2$ weighting is demonstrated in knee and liver studies. Because $Q^2MRI$ has no requirements with regards to the imaging protocol used for input image acquisition, a variety of quantitative parametric maps can be derived from images that are routinely obtained in clinical practice, facilitating the conduct of quantitative image analysis in prospective and retrospective studies. In radiation therapy, $Q^2MRI$ will aid MRI-only treatment planning [54] and image-guided therapy [55]. With the application of Bloch equations, MR imaging can be tailored to individual patients, with contrasts chosen to meet their specific clinical demands. The proposed $Q^2MRI$ method can be further extended to quantification of other tissue parameters for diverse clinical applications. As healthcare and biomedicine are transforming into digitized and data intensive disciplines, quantitative imaging capabilities enabled by $Q^2MRI$ may find valuable applications in the future and promise to benefit a wide spectrum of biomedical applications.

**Acknowledgments:** The authors would like to thank Drs. Charles Mistretta, Brian Rutt, Debiao Li, Rohan Dharmakumar, Garry Gold, Zhitao Li, Christopher Sandino, Aiming Lu, Huimin Wu, Jing Liu, Nicole Le, Maria Chan, and Cheng Tang for their helpful discussion.

**Funding:** This research is partially supported by NIH/NCI (1R01 CA256890, 1R01 CA227713), NIH/NIAMS (1R01 AR068987), and NIH/NINDS (1R01 NS092650), and NIH/NIBIB (1R01 EB026136). The contents of this article are solely the responsibility of the authors and do not necessarily represent the official NIH views.

**Author contributions:** L.X. proposed the strategy of using deep learning to separate the influence of different contributive factors from single MR images and thus derive quantitative parametric maps from clinical routine without additional acquisition. He guided the direction of the research. Y.W. conducted parametric mapping experiments, hypothesized how deep learning extracts *a priori* information and exploits the interdependency within/across MRI parameter subspaces. Y.M. and J.D. provided knee data and proposed to exclude $B_1$ map in $T_1$ mapping. S.V., Y.K., M.A. and J.P. provided liver data and suggested to use complex data for parametric mapping. S.V., E.Y.C. and S.H. evaluated the clinical efficacy. N.K., D. C. and H.R. performed preprocessing. The manuscript was written by Y. W., L.X., S.V. and J.D., and revised by all authors.

**Competing interests:** L.X. and Y.W. are co-inventors on patents based on this work [66, 67].

**Data and materials availability:** Source code is available from the corresponding author upon request.

# Supplementary Material

## A.1. Image acquisition

In this study, $T_1$- and $T_2^*$-weighted knee images were acquired on 3T GE MR750 scanners [39, 40]. For every subject, the following images were obtained: four $T_1$-weighted images acquired using an ultrashort echo time (UTE) cones sequence with flip angles of 5°, 10°, 20°, and 30° respectively, a TE (echo time) of 32 μs, and a TR (time of repetition) of 20 ms; six $T_2^*$-weighted images acquired using a multi-echo sequence with TEs of 0.032, 4.4, 8.8, 13.2, 17.6 and 22 ms, a TR of 28 ms, and a flip angle of 16°; and finally a $B_1$ map measured using an actual flip angle sequence [9] with a TE of 32 μs, two interleaved TRs (20/100ms), and a flip angle of 45°. Notice that the incorporation of the $B_1$ map was important for variable-flip angle $T_1$ mapping (a 20% difference in $B_1$ map, which is typical at 3T, causes a 44% error in $T_1$ map). The ground truth $T_1$ and $T_2^*$ maps were extracted using the Levenberg-Marquardt algorithm [56].

In the liver, free-breathing abdominal MR images were acquired on three 3T GE MR750 scanners [44]. For every subject, six $T_2^*$-weighted images were acquired using a multi-echo cones sequence with TEs of 0.036, 1.048, 2.060, 3.072, 4.084, and 5.096 ms respectively, a TR of 11ms, and a flip angle of 3°. The ground truth $R_2^*$, $B_0$, and $PDFF$ maps were calculated from six multi-echo images based on the confounder model, which combines least squares fitting of complex-valued source images with multi-peak fat modeling. While cones acquisition is robust to motion, some data acquired in the free-breathing condition was of low image quality and thus excluded. High quality images from 26 patients were used in this study.

## A.2. Segmentation

In preprocessing, the torso is first delineated to eliminate background noise from outside the human body. For regional quantitative analysis, the whole liver was segmented using the region growing algorithm (S1). The generated ROI was then applied on predicted and ground truth parametric maps for quantitative analysis.

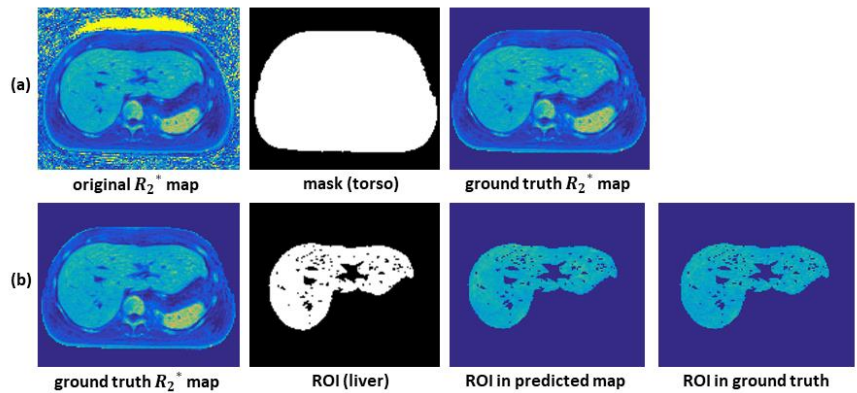

**S1**. Segmentation as preprocessing. (a) Torso is delineated to prevent background noise from outside the human body. (b) The whole liver is segmented using the region growing algorithm.

## A.3. Hyperparameters of deep neural networks

For the proposed parametric mapping tasks, a self-attention convolutional neural network architecture was employed (S2). The network has a unique shortcut connection pattern as inspired by V-net, DenseNet, and other methods [57-61]. Here, relatively dense local shortcuts forward feature maps to all the convolutional blocks in the same hierarchical level, and global shortcuts connect the encoder path and the decoder path, effectively facilitating residual learning [62, 63].

In the hierarchical network, features are extracted at various scales. At each level, there are three convolutional blocks. Every convolutional block is composed of a convolutional layer, a self-attention layer, and a nonlinear activation layer. At the convolutional layer, image features are extracted using 3×3 convolutional kernels. At the self-attention layer, the self-attention maps are derived for feature maps extracted at the preceding convolutional layer. At the activation layer, the Parametric Rectified Linear Unit (PReLU) function is applied. Down-sampling and up-sampling are accomplished using 2×2 convolutional kernels with a stride of 2 [59]. In the knee studies, every network had five hierarchical levels (with 16, 32, 64, 128, and 256 channels in every level, respectively). In the liver study, the networks for $R_2^*$ mapping and $B_0$ mapping had three hierarchical levels (with 16, 32, and 64 channels in every level, respectively), whereas the network for $PDFF$ mapping had five hierarchical levels (with 16, 32, 64, 128, and 256 channels in every level, respectively).

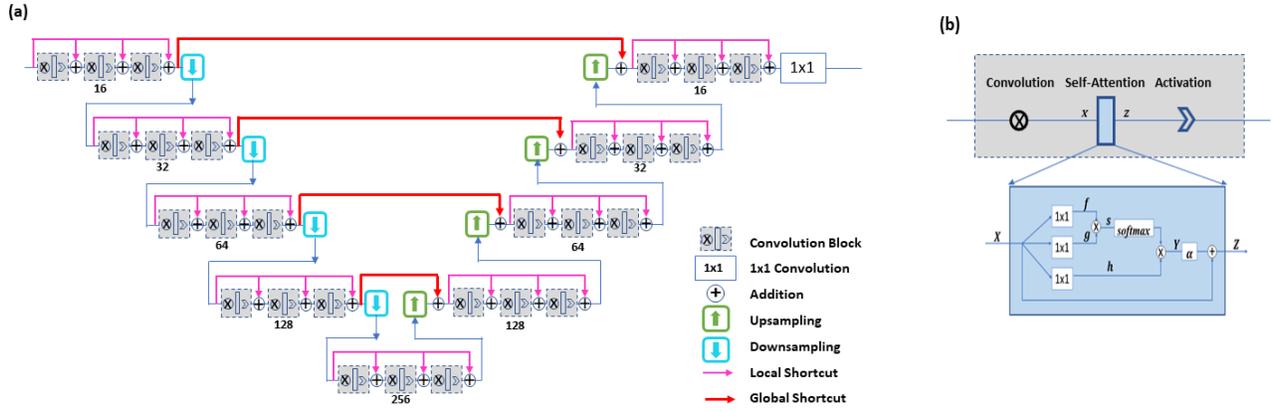

**S2.** The proposed self-attention convolutional neural network. (a) A hierarchical network with global shortcuts and densely connected local shortcuts. (b) A convolutional block with a self-attention layer integrated. X is the input (preceding feature map), Y is the attention map, and Z is the output of the convolutional block. The attention map Y is determined by s (relevance between the given pixel and the other pixel) and h (feature representation of the other pixel). In the output Z, the contributions from local information (feature map X) and non-local information (attention map Y) are balanced by a scalar $\alpha$.

Motivated by [33-37], the attention mechanism has been integrated into the deep neural network to exploit image-wide/non-local *a priori* information with higher efficiency. In conventional deep neural networks, long-range dependency (i.e., interaction between widely separated pixels) is progressively propagated across multiple layers, which may cause optimization difficulties. A self-attention network, on the other hand, overcomes this limitation by establishing direct interactions between a given pixel and all the other pixels in the feature map. The intensity of a direct interaction is determined by two factors: relevance between the given pixel and the other pixel, and features of the other pixel. We quantify the relevance between the two pixels using an embedded Gaussian function $s(X_i, X_j) = exp\{(W_f X_i)'(W_g X_j)\}$, and represent the feature of the other pixel with a linear function $h(X_j) = W_h X_j$, where $W_f$, $W_g$, and $W_h$ are weight matrices learned in training. Unlike a transformer [36], the proposed network combines a self-attention module with the convolutional layer, where the former extracts non-local knowledge and the latter acquires local information. The attention mechanism is more influential when the receptive field of the network is small. In the liver study (where the network had fewer hierarchical levels), $R_2^*$ mapping could not be accomplished without self-attention.

In the loss function $loss = l_1 + \lambda * (1 - SSIM)$, SSIM is defined as $SSIM(x,y) = \frac{(2\mu_x\mu_y + C_1)(2\sigma_x\sigma_y + C_2)}{(\mu_x^2 + \mu_y^2 + C_1)(\sigma_x^2 + \sigma_y^2 + C_2)}$, where $\mu_x$, $\mu_y$, $\sigma_x$, and $\sigma_y$ correspond to the mean and standard deviation of signal intensity in the reconstructed image $x$ and the ground truth $y$, and $C_1$ and $C_2$ are constants [42]. Comparison of the mixed $l_1$-$l_{SSIM}$ loss with $l_1$ loss has been performed previously in [26]. To evaluate prediction accuracy, correlation coefficient was used, defined as $orr(x,y) = \frac{2\sigma_{xy}}{\sigma_x^2 + \sigma_y^2 + (\mu_x - \mu_y)^2}$.

The network was implemented on a TensorFlow-based AI platform NiftyNet [64, 65]. Computations were performed on a desktop computer running Linux operating system with an Intel i77700K CPU (4.2 GHz, and 32GB memory) and Nvidia GPU GeForce GTX1070.

### A.4. Retrospective contrast tuning

After quantitative parametric maps are estimated using $Q^2MRI$, MRI contrast can be retrospectively tuned with the application of Bloch equations. As an example, we derived variable-flip-angle images from a single $T_1$-weighted image. Based on the $T_1$, $\rho$ and $B_1$ maps predicted from a $T_1$-weighted image (acquired using a flip angle of 30°), other $T_1$-weighted images were obtained according to $S = \rho \cdot \sin(\alpha) \frac{1 - e^{-TR/T_1}}{1 - \cos(\alpha) \cdot e^{-TR/T_1}}$. Here, $B_1$ inhomogeneity is taken into consideration via $\alpha = \alpha_{nominal} \cdot B_1$, where $\alpha_{nominal}$ is the nominal flip angle specified by the imaging protocol and $\alpha$ is the actual flip angle that takes effects. The predicted images (corresponding to flip angles of 5°, 10°, and 20°, respectively) were compared with the ground truth, and high fidelity was achieved (S3, Table 2). More tissue contrasts can be obtained by taking other values of imaging parameters (flip angle, TR).

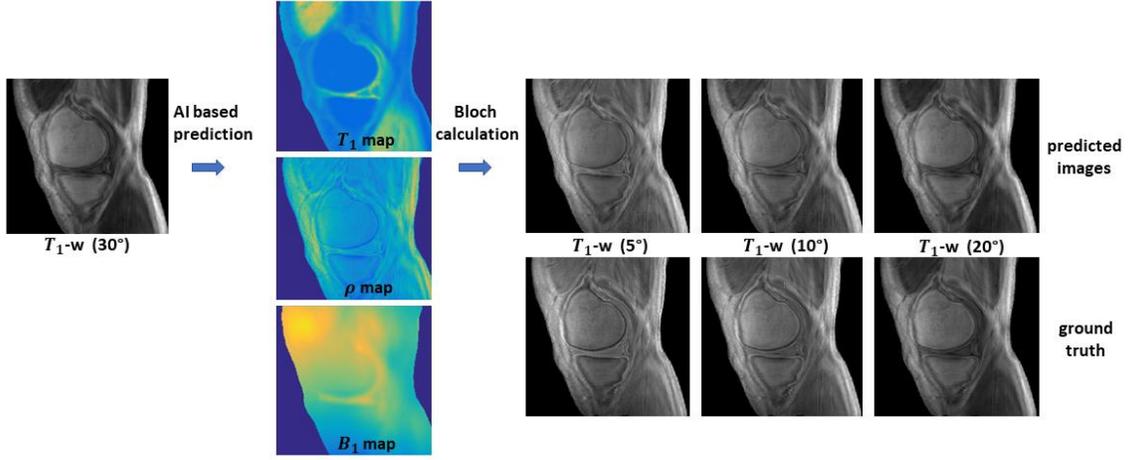

**S3.** Retrospective tuning of tissue contrast in $T_1$-weighted image. Given a single $T_1$-weighted image (acquired using a flip angle of 30°), $T_1$, $\rho$ and $B_1$ maps are predicted and used to synthesize images (corresponding to flip angles of 5°, 10°, and 20°). High fidelity is achieved in the resultant images.

| input \ target | Correlation $T_1$ w (5°) | $T_1$ w (10°) | $T_1$ w (20°) | $T_1$ w (30°) | input \ target | $l_1$ Error $T_1$ w (5°) | $T_1$ w (10°) | $T_1$ w (20°) | $T_1$ w (30°) |
|---|---|---|---|---|---|---|---|---|---|
| $T_1$ w (5°) | 0.9880 ± 0.0044 | 0.9774 ± 0.0061 | 0.9849 ± 0.0040 | 0.9762 ± 0.0072 | $T_1$ w (5°) | 0.0585 ± 0.0111 | 0.0793 ± 0.0137 | 0.0693 ± 0.0156 | 0.0892 ± 0.0216 |
| $T_1$ w (10°) | 0.9817 ± 0.0054 | 0.9933 ± 0.0018 | 0.9915 ± 0.0009 | 0.9845 ± 0.0034 | $T_1$ w (10°) | 0.0716 ± 0.0130 | 0.0441 ± 0.0069 | 0.0537 ± 0.0085 | 0.0730 ± 0.0152 |
| $T_1$ w (20°) | 0.9754 ± 0.0086 | 0.9888 ± 0.0030 | 0.9950 ± 0.0009 | 0.9893 ± 0.0028 | $T_1$ w (20°) | 0.0819 ± 0.0171 | 0.0570 ± 0.0093 | 0.0426 ± 0.0065 | 0.0617 ± 0.0120 |
| $T_1$ w (30°) | 0.9702 ± 0.0104 | 0.9853 ± 0.0051 | 0.9929 ± 0.0009 | 0.9912 ± 0.0021 | $T_1$ w (30°) | 0.0897 ± 0.0198 | 0.0661 ± 0.0129 | 0.0504 ± 0.0083 | 0.0543 ± 0.0116 |

| | | Predicting images from a $T_1$ w image | | | | Predicting images from $T_1$ w and $T_2$ w images | | | |
|---|---|---|---|---|---|---|---|---|---|
| | Output \ Input | $T_1$ w (5°) | $T_1$ w (10°) | $T_1$ w (20°) | $T_1$ w (30°) | $T_1$ w (5°) | $T_1$ w (10°) | $T_1$ w (20°) | $T_1$ w (30°) |
| $l_1$ Error | 5° | 0.0559 ± 0.0104 | 0.0706 ± 0.0063 | 0.1177 ± 0.0174 | 0.1472 ± 0.0212 | 0.0719 ± 0.0118 | 0.0688 ± 0.0087 | 0.0923 ± 0.0130 | 0.1159 ± 0.0144 |
| | 10° | 0.0777 ± 0.0141 | 0.0456 ± 0.0046 | 0.0918 ± 0.0145 | 0.1256 ± 0.0213 | 0.0714 ± 0.0140 | 0.0368 ± 0.0048 | 0.0760 ± 0.0096 | 0.1071 ± 0.0136 |
| | 20° | 0.1040 ± 0.0114 | 0.0765 ± 0.0067 | 0.0743 ± 0.0184 | 0.1093 ± 0.0250 | 0.0826 ± 0.0115 | 0.0570  0.0064 | 0.0679  0.0165 | 0.1044 ± 0.0244 |
| | 30° | 0.1106 ± 0.0111 | 0.0900 ± 0.0090 | 0.0843 ± 0.0133 | 0.0916 ± 0.0198 | 0.0830 ± 0.0111 | 0.0646  0.0078 | 0.0733  0.0100 | 0.0887 ± 0.0177 |
| SSIM | 5° | 0.9909 ± 0.0017 | 0.8894 ± 0.0288 | 0.8217 ± 0.0341 | 0.7812 ± 0.0363 | 0.9540 ± 0.0057 | 0.8879 ± 0.0293 | 0.8433 ± 0.0357 | 0.8135 ± 0.0381 |
| | 10° | 0.8897 ± 0.0279 | 0.9849 ± 0.0029 | 0.8930 ± 0.0212 | 0.8525 ± 0.0247 | 0.8985 ± 0.0278 | 0.9930 ± 0.0024 | 0.9089 ± 0.0224 | 0.8743 ± 0.0250 |
| | 20° | 0.8220 ± 0.0348 | 0.8910 ± 0.0230 | 0.9666 ± 0.0077 | 0.9026 ± 0.0120 | 0.8681 ± 0.0331 | 0.9194  0.0221 | 0.9646  0.0083 | 0.9018 ± 0.0136 |
| | 30° | 0.7812 ± 0.0373 | 0.8501 ± 0.0270 | 0.9071 ± 0.0111 | 0.9596 ± 0.0085 | 0.8497 ± 0.0356 | 0.8969  0.0251 | 0.9127  0.0111 | 0.9412 ± 0.0076 |
| Correlation Coefficient | 5° | 0.9977 ± 0.0008 | 0.9919 ± 0.0027 | 0.9796 ± 0.0073 | 0.9718 ± 0.0103 | 0.9961 ± 0.0013 | 0.9940 ± 0.0024 | 0.9889 ± 0.0039 | 0.9843 ± 0.0055 |
| | 10° | 0.9918 ± 0.0026 | 0.9969 ± 0.0011 | 0.9890 ± 0.0035 | 0.9822 ± 0.0058 | 0.9939 ± 0.0021 | 0.9983 ± 0.0008 | 0.9929 ± 0.0024 | 0.9877 ± 0.0043 |
| | 20° | 0.9821 ± 0.0049 | 0.9906 ± 0.0028 | 0.9929 ± 0.0034 | 0.9873 ± 0.0051 | 0.9900 ± 0.0033 | 0.9949 ± 0.0018 | 0.9949 ± 0.0017 | 0.9895 ± 0.0029 |
| | 30° | 0.9786 ± 0.0060 | 0.9870 ± 0.0038 | 0.9909 ± 0.0033 | 0.9899 ± 0.0044 | 0.9888 ± 0.0037 | 0.9932 ± 0.0024 | 0.9933 ± 0.0020 | 0.9914 ± 0.0033 |

**Table 2**. Evaluation of variable contrast image predictions. In each row, images with a certain contrast (presumably acquired using a specific flip angle) are predicted from different input images. Both low $l_1$ errors (between 0.04 and 0.09) and high correlation coefficients (ranging from 0.97 to 0.99) are consistently achieved on a pixel basis when different flip angles are used for input and output images.

## A.5. Interdependency between various relaxation parametric maps

In quantum statistics [48], $T_1$, $T_2$, and $\rho$ are related by the following:
$\frac{1}{T_1} = \frac{3}{10} b^2 [J(\omega_0) + 4J(2\omega_0)]$ and $\frac{1}{T_2} = \frac{3}{20} b^2 [3J(0) + 5J(\omega_0) + 2J(2\omega_0)]$,

where $J(\omega_0)$ is the spectral density function at the resonance frequency expressed by:
$J(\omega_0) = \int_{-\infty}^{\infty} G(\tau) exp(-i\omega_0 \tau) \, d\tau. \; b = -\frac{h\gamma^2}{r^3}$.

In $Q^2 MRI$, this complicated relationship between various tissue relaxation properties is exploited and encoded into a parametric mapping model as *a priori* knowledge. Along with other spatial and statistical information, the correlation between various parametric maps is deciphered.

Integrating the correlation between various parametric maps into quantitative parametric mapping model enables the derivation of $T_1$ map from a single $T_1$-weighted image. In a $T_1$-weighted image, the signal intensity is a function of both $T_1$ and $\rho$, $S = f_1(T_1, \rho)$. If $\rho$ can be represented as a function of $T_1$ (i.e., $\rho = m(T_1)$), then the signal intensity is a function of $T_1$ (i.e., $S = f_2(T_1)$). The inverse problem of deriving $T_1$ from a single $T_1$-weighted image is no longer an ill-posed problem.

## A.6. Exploiting *a priori* information in high-dimensional MRI parameter space

In $Q^2MRI$, *a priori* information is exploited not only within the tissue property subspace (A.5), but also across different subspaces. For instance, a subtle connection between the tissue property subspace and the system condition subspace is established when the interdependency between $B_1$ and MR signals are exploited.

It is established that there is a relationship between $B_1$ and MR signal caused by the electrodynamic interaction between the incident transmission radiofrequency field and human anatomy [20]. In $Q^2MRI$, this relationship is exploited as *a priori* knowledge and encoded into parametric mapping models.

Moreover, in $Q^2MRI$, spatial and statistical knowledge is seamlessly incorporated. In previous approaches, similar information (e.g., the statistical distribution of $T_1$ in different tissues [54, 55] or the linear relationship between $1/\rho$ and $1/T1$ in brain tissue [56]) were used to estimate the $B_1$ map from multiple input images, but it was difficult to combine them with each other or with physical models. In $Q^2MRI$, with seamless incorporation of *a priori* information relevant to different aspects, the $B_1$ map is estimated from only a single input image.

The implicit integration of the predicted $B_1$ map into the $T_1$ mapping model helps compensate for $B_1$ inhomogeneity without the need to perform actual measurement of the $B_1$ map. In a $T_1$-weighted image, the signal intensity is a function of $T_1$, $\rho$, and $B_1$ ($S = f_3(T_1, \rho, B_1)$). If $B_1$ can be represented as a function of the signal intensity ($B_1 = \text{n}(S)$) and $\rho = \text{m}(T_1)$, then the $T_1$ map can be derived from a single $T_1$-weighted image with $B_1$ compensation being automatically achieved ($T_1 = f_4(S)$). Here, the acquisition parameter subspace and the system condition subspace are connected ($\alpha = \alpha_{nominal} \cdot B_1$).